\documentclass[pdflatex,sn-mathphys-num]{sn-jnl}

\usepackage{graphicx}%
\graphicspath{{./figures/}}
\usepackage{multirow}%
\usepackage{float}
\usepackage{subcaption}
\usepackage{amsmath,amssymb,amsfonts}%
\usepackage{booktabs}
\usepackage[table]{xcolor}
\usepackage{graphicx} 
\usepackage{amsthm}%
\usepackage{mathrsfs}%
\usepackage[title]{appendix}%
\usepackage{textcomp}%
\usepackage{manyfoot}%
\usepackage{algorithm}%
\usepackage{algorithmicx}%
\usepackage{algpseudocode}%
\usepackage{listings}%
\usepackage{natbib}
\usepackage{url}
\usepackage{cleveref}
\usepackage{pgfplots}
\pgfplotsset{compat=1.18}
\usepackage{pdflscape} 
\usepackage{multirow}
\usepackage{tabularx}
\usepackage{array}
\usepackage{tikz}
\usepackage{orcidlink}
\usetikzlibrary{positioning, shapes, arrows.meta}

\newcolumntype{P}[1]{>{\raggedright\arraybackslash}p{#1}}  
\newcolumntype{L}{>{\raggedright\arraybackslash}X}         
\newcolumntype{T}{>{\ttfamily\raggedright\arraybackslash}X}

\newcommand{\best}[1]{\textbf{\textcolor{green!50!black}{#1}}}
\newcommand{\worst}[1]{\textcolor{red}{\underline{#1}}}

\def\BibTeX{{\rm B\kern-.05em{\sc i\kern-.025em b}\kern-.08em
T\kern-.1667em\lower.7ex\hbox{E}\kern-.125emX}}
\usepackage[acronym,toc]{glossaries}
\newacronym{eg}{EG}{Example of Acronym}
\newacronym{snn}{SNN}{Spiking Neural Network}
\newacronym{dl}{DL}{Deep Learning}
\newacronym{nlp}{NLP}{Natural Language Processing}
\newacronym{cnn}{CNN}{Convolutional Neural Network}
\newacronym{stdp}{STDP}{Spike timing-dependent plasticity}
\newacronym{ann}{ANN}{artificial neural network}
\newacronym{resume}{ReSuMe}{Remote Supervised Method}
\newacronym{hh}{HH}{Hodgkin-Huxley}
\newacronym{iaf}{IAF}{Integrate and Fire}
\newacronym{lif}{LIF}{Leaky Integrate-and-Fire}
\newacronym{izk}{IZK}{Izhikevich}
\newacronym{nest}{NEST}{NEural Simulation Tool}
\newacronym{lsb}{LSB}{least significant bit}
\newacronym{dct}{DCT}{Discrete Cosine Transform}
\newacronym{dft}{DFT}{Discrete Fourier Transform}
\newacronym{isic}{ISIC}{International Skin Imaging Collaboration}
\newacronym{dog}{DOG}{Difference of Gaussians}
\newacronym{bcs}{BCS}{British Computing Society}
\newacronym{dpa}{DPA}{Data Protection Act}
\newacronym{ai}{AI}{Artificial Intelligence}
\newacronym{ram}{RAM}{random access memory}
\newacronym{kb}{KB}{kilobytes}
\newacronym{mb}{MB}{megabytes}
\newacronym{crisp-dm}{CRISP-DM}{cross-industry process for data mining}
\newacronym{ml}{ML}{machine learning}
\newacronym{iot}{IoT}{internet of things}
\newacronym{gdpr}{GDPR}{general data protection regulation}
\newacronym{sipi}{SIPI}{Signal and Image Processing Institute}
\newacronym{psnr}{PSNR}{peak signal-to-noise ratio}
\newacronym{ssim}{SSIM}{structural similarity index}
\newacronym{rmse}{RMSE}{root mean squared error}
\newacronym{rgba}{RGBA}{red, green, blue, alpha}
\newacronym{axi}{AXI}{Advanced eXtensible Interface}
\newacronym{dma}{DMA}{Direct Memory Access}
\newacronym{fpga}{FPGA}{field-programmable gate array}
\newacronym{spa}{SPA}{Sample Pair Analysis}
\newacronym{ws}{WS}{Weighted Stego-image Residuals}
\newacronym{bpp}{bpp}{bits per pixel}
\usepackage{placeins}

\theoremstyle{thmstyleone}%
\theoremstyle{thmstyletwo}%
\newtheorem{example}{Example}%

\theoremstyle{thmstylethree}%

\crefname{figure}{Fig.}{Figs.}

\begin{document}

\title[Article Title]{SteganoSNN: SNN-Based Audio-in-Image Steganography with Encryption}

\author*[1]{\fnm{Biswajit Kumar} \sur{Sahoo}\orcidlink{0009-0000-2045-9765}}\email{21ec01053@iitbbs.ac.in}
\equalcont{These authors contributed equally to this work.}

\author*[2]{\fnm{Pedro} \sur{Machado}\orcidlink{0000-0003-1760-3871}}\email{pedro.machado@ntu.ac.uk}
\equalcont{These authors contributed equally to this work.}

\author[2]{\fnm{Andreas} \sur{Oikonomou}\orcidlink{0000-0002-5069-3971}}\email{andreas.oikonomou}

\author[2]{\fnm{Isibor Kennedy}\sur{Ihianle}\orcidlink{0000-0001-7445-8573}}\email{isibor.ihianle@ntu.ac.uk}

\author[1]{\fnm{Srinivas} \sur{Boppu}\orcidlink{0000-0001-9028-2563}}\email{srinivas@iitbbs.ac.in}

\affil*[1]{\orgdiv{School of Electrical and Computer Sciences}, \orgname{Indian Institute of Technology Bhubaneswar}, \orgaddress{\street{Argul}, \city{Khordha}, \postcode{752050}, \state{Odisha}, \country{India}}}

\affil[2]{\orgdiv{Computer Sciences}, \orgname{Nottingham Trent University}, \orgaddress{\street{Clifton Campus}, \city{Nottingham}, \postcode{NG11 8NS}, \state{Nottinghamshire}, \country{UK}}}

\abstract{Secure data hiding remains a fundamental challenge in digital communication, requiring a careful balance between computational efficiency and perceptual transparency. The balance between security and performance is increasingly fragile with the emergence of generative \gls*{ai} systems capable of autonomously generating and optimising sophisticated cryptanalysis and steganalysis algorithms, thereby accelerating the exposure of vulnerabilities in conventional data-hiding schemes.

This work introduces \textbf{SteganoSNN}, a neuromorphic steganographic framework that exploits \glspl*{snn} to achieve secure, low-power, and high-capacity multimedia data hiding. Digitised audio samples are converted into spike trains using \gls*{lif} neurons, encrypted via a modulo-based mapping scheme, and embedded into the least significant bits of \gls*{rgba} image channels using a dithering mechanism to minimise perceptual distortion. Implemented in Python using NEST and realised on a PYNQ-Z2 \gls*{fpga}, \textbf{SteganoSNN} attains real-time operation with an embedding capacity of 8~\gls*{bpp}. Experimental evaluations on the DIV2K~2017 dataset demonstrate image fidelity between 40.4~dB and 41.35~dB in \gls*{psnr} and \gls*{ssim} values consistently above 0.97, surpassing SteganoGAN in computational efficiency and robustness. \textbf{SteganoSNN} establishes a foundation for neuromorphic steganography, enabling secure, energy-efficient communication for Edge-\gls*{ai}, \gls*{iot}, and biomedical applications.
}

\keywords{Spiking Neural Networks, Steganography, Leaky Integrate-and-Fire, FPGA, Neuromorphic Computing, Edge AI, Image Processing, Audio Hiding}



\maketitle

\section{Introduction}\label{sec1}

\glspl*{snn} have recently attracted significant attention for their biologically inspired and computationally efficient design, finding applications in areas such as image classification~\cite{9103079} and object detection. Unlike traditional \glspl*{ann}, which rely on continuous-valued activations, \glspl*{snn} transmit information through discrete spike events that more closely emulate the signalling mechanisms of biological neurons. The event-driven representation allows \glspl*{snn} to process spatio-temporal information efficiently, making them particularly valuable for time-sensitive tasks such as dynamic pattern recognition, neuromorphic sensing, and real-time decision-making. Despite these advantages, the use of \glspl*{snn} in the field of steganography remains unexplored.

Image steganography involves concealing information within a cover image in such a way that the presence of the hidden content is imperceptible to the human visual system. Steganography contrasts with cryptography, in which the existence of secret data is apparent but the content remains unreadable without a key. Steganography has a wide range of legitimate applications, including secure military communication, protection of patient privacy in medical data transmission, copyright authentication, and confidential document exchange~\cite{MANDAL20221451}. However, as with other data-hiding methods, it also carries the potential for misuse, such as embedding malicious payloads or distributing covert information~\cite{9599677}. It is therefore important to advance steganographic methods in a way that balances capability, detectability, and responsible use.

As far as the authors are aware, \glspl*{snn} have not yet been explored for steganographic purposes. This represents a clear gap in the literature and an opportunity to investigate their potential benefits in this domain. Existing steganographic techniques such as those based on the \gls*{dct}, wavelet transform, or adaptive machine-learning-based embedding can be computationally intensive, often limiting their deployment on devices with restricted processing power or energy budgets. In contrast, \glspl*{snn} offer an inherently efficient computational model that can operate asynchronously and with sparse activity, making them highly suitable for lightweight or real-time embedding systems. 

Furthermore, traditional methods such as \gls*{lsb} substitution or \gls*{dct}-based embedding are known to be vulnerable to compression, scaling, and noise, leading to degradation or loss of the hidden information. By exploiting the temporal dynamics and resilience of spike-based encoding, an \gls*{snn}-driven steganographic system may provide greater robustness against such distortions while maintaining low computational overhead.

This article contributes to the existing body of knowledge in both \gls*{snn} modelling and steganography by proposing a novel framework that conceals audio data within a cover image using \gls*{snn}-derived spike patterns as an intermediate encoding mechanism. The proposed system, termed \emph{SteganoSNN}, integrates a biologically inspired spiking model with a compact encryption scheme and an efficient \gls*{lsb}-based embedding strategy to achieve high payload capacity, strong perceptual fidelity, and low detectability. From a practical standpoint, the computational efficiency of \glspl*{snn} makes the proposed approach well-suited for applications where hardware constraints are critical. For example, data from medical devices such as pacemakers, insulin pumps, and portable diagnostic sensors could be securely embedded within routine communication signals, ensuring privacy with negligible performance impact. Although this study focuses on modest network sizes due to hardware limitations, the underlying concept is scalable to more complex architectures and larger data volumes.

The remainder of this article is structured as follows: Section~\ref{sec2:lr} reviews the related literature on \glspl*{snn} and image steganography. Section~\ref{sec:methodology} presents the proposed methodology. Section~\ref{sec:results} discusses the implementation, evaluation results, and steganalysis robustness. Finally, conclusions and potential directions for future research are drawn in Section~\ref{sec:results}.

\section{Related Work}\label{sec2:lr}
\glspl*{snn} mimic biological neural networks through neurons interconnected by synapses~\cite{9787485}, encoding information as discrete spikes. These spikes emulate the action potentials observed in biological neurons~\cite{malcolm2023comprehensive} and allow the temporal structure of signals to be represented directly. Owing to this event-driven nature, \glspl*{snn} are well suited for time-series~\cite{lv2024efficient} and event-based data~\cite{Stewart_2022}, where temporal precision and low-latency inference are crucial. Compared with conventional \glspl*{ann}, \glspl*{snn} can achieve significantly lower power consumption for equivalent computational workloads, offering a path toward more sustainable and energy-efficient \gls*{dl} systems~\cite{shi2024towards,venzke2020artificial,Plagwitz2024}. This makes them particularly attractive for edge computing and real-time applications in which devices operate on limited or battery-powered hardware.

Neuronal dynamics in \glspl*{snn} can be modelled at various levels of biological realism. The pioneering \gls*{hh} model~\cite{Hodgkin1952} provides the most detailed representation of membrane currents and ion-channel kinetics but is computationally expensive. The simplistic \gls*{iaf} model~\cite{burkitt} abstracts these processes by accumulating weighted synaptic inputs until a threshold is reached. The \gls*{lif} model~\cite{Gerstner_Kistler_2002,lu2022linear} extends this formulation by introducing a passive “leak” term, causing the membrane potential to decay toward its resting state when no inputs are received. A further compromise between realism and efficiency is offered by the \gls*{izk} model~\cite{1257420,1333071}, which reproduces diverse neuronal firing patterns, including all known cortical neuron types, with minimal computational overhead. Model selection typically depends on the intended balance between fidelity, efficiency, and scalability. In this study, the \gls*{lif} model is adopted owing to its stability and reproducibility under constant input, making it well suited for the controlled spike pattern generation required by our proposed framework.

To implement and simulate such models, software tools such as NEST and Brian~2 are widely used. NEST~\cite{Gewaltig:2007,nest_iaf_psc_alpha} provides efficient event-driven simulation of large spiking networks and supports Python interfaces for flexible configuration. Brian~2~\cite{10.7554/eLife.47314} offers an intuitive Python-based framework with automatic code generation for different back ends. In this work, we use NEST to generate reproducible spike patterns from the \gls*{lif} neuron model under constant input current, which later form the temporal basis of our encoding scheme.

Beyond software simulation, there is increasing interest in deploying \glspl*{snn} on hardware accelerators, especially \glspl*{fpga}. \glspl*{fpga} combine reconfigurability with high parallelism and deterministic timing, providing an ideal platform for energy-efficient neuromorphic computation~\cite{Plagwitz2024}. Implementations such as spike-based convolutional networks, event-driven controllers, and hybrid \gls*{cnn}–\gls*{snn} systems have demonstrated orders-of-magnitude reductions in power consumption compared to CPU or GPU inference~\cite{kim2020,lee2020spike,9340861}. Such architectures are particularly advantageous in embedded and real-time applications, where latency and energy efficiency are critical. The framework proposed in this paper aligns with these objectives, as it can be readily realised on \gls*{fpga}-based platforms such as PYNQ-Z2 through lightweight Verilog IP cores for spike encoding, encryption, and embedding.

Parallel to advances in neuromorphic computing, steganography has evolved as an essential tool for information hiding and secure data communication. It aims to conceal the presence of a secret payload within a cover medium such that the alteration remains visually imperceptible~\cite{subramanian2021image,MANDAL20221451}. This differs from cryptography, which ensures confidentiality but not invisibility, since encrypted data still indicate the presence of a secret message. Steganography supports diverse legitimate applications including copyright protection, military communication, and medical data privacy~\cite{MANDAL20221451}, though it can also be misused for malicious purposes~\cite{9599677}, underscoring the importance of developing methods that balance capacity, fidelity, and detectability.

Among spatial-domain techniques, \gls*{lsb} substitution is the most widely used due to its simplicity and high payload capacity~\cite{solak2019image,cisska-lsb}. Each pixel in an \gls*{rgba} image typically consists of four 8-bit channels, allowing the two or more least significant bits of each channel to carry hidden data. Although \gls*{lsb}-based schemes are easy to implement and computationally efficient, they are vulnerable to compression, noise, and advanced steganalysis attacks~\cite{subramanian2021image,8577655,KARAMPIDIS2018217}. Transform-domain approaches, such as those employing the \gls*{dct}, \gls*{dft}, or wavelet decomposition, offer improved robustness by embedding information into frequency coefficients rather than pixel values~\cite{9187785}. However, these methods incur greater computational complexity, making them less suitable for low-power or real-time applications.

This article proposes to bridge these domains by introducing an \emph{\gls*{snn}-inspired steganographic framework} in which spike-time patterns generated by a \gls*{lif} neuron are used as an intermediate encoding mechanism before embedding into \gls*{rgba} images. By combining the temporal robustness of spike-based representations with the efficiency of spatial-domain embedding, the proposed system achieves high payload capacity and perceptual fidelity while maintaining compatibility with FPGA-based deployment for energy-efficient, real-time operation.

\section{Methodology}\label{sec:methodology}

\subsection{Overview}
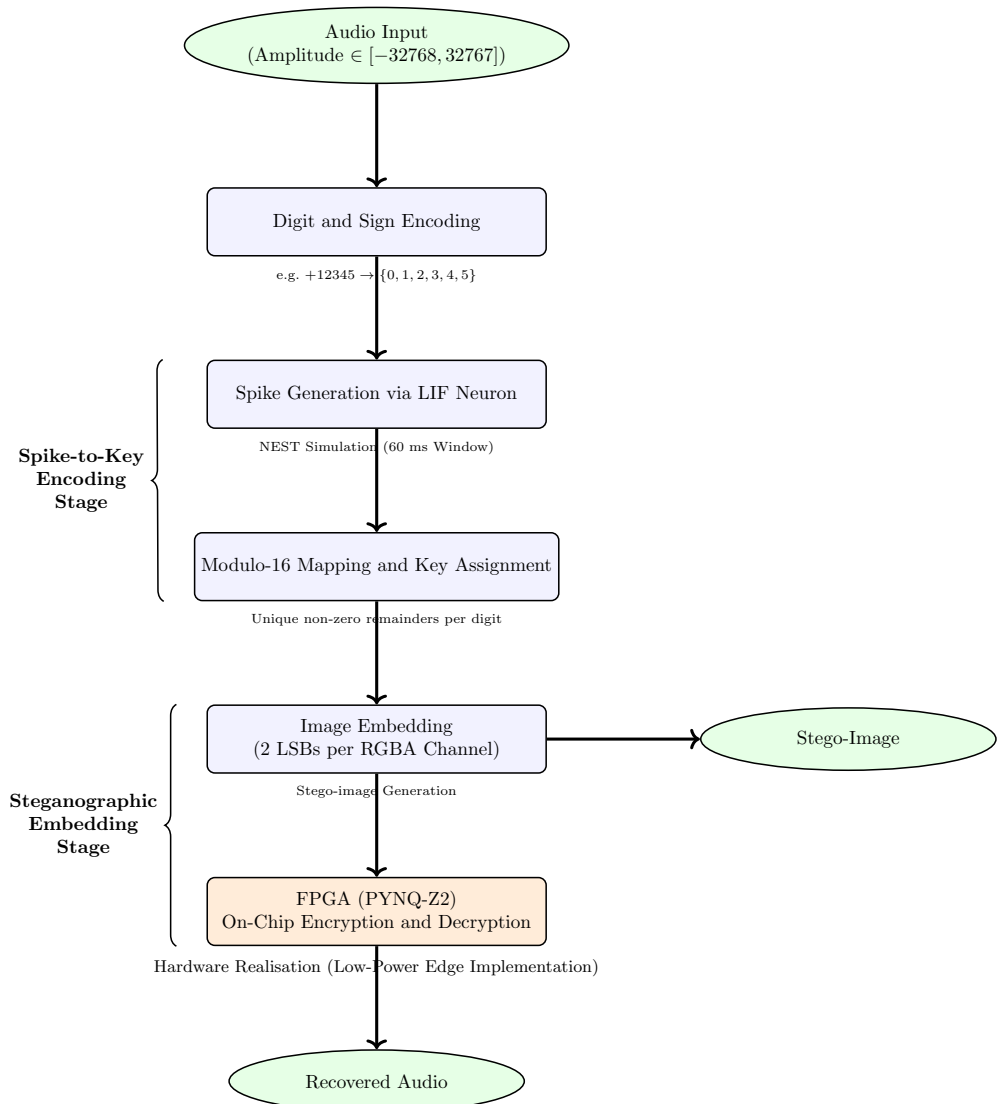
\begin{figure*}[]
\centering
\resizebox{\textwidth}{!}{%
\begin{tikzpicture}[
    node distance=1.9cm,
    box/.style={rectangle, draw=black, thick, rounded corners, align=center,
                minimum width=6.2cm, minimum height=1.25cm, fill=blue!5, font=\normalsize},
    hw/.style={rectangle, draw=black, thick, rounded corners, align=center,
               minimum width=6.2cm, minimum height=1.25cm, fill=orange!15, font=\normalsize},
    cloud/.style={ellipse, draw=black, thick, align=center, fill=green!10,
                  minimum height=1.15cm, minimum width=5.4cm, font=\normalsize},
    arrow/.style={->, ultra thick},
    dashedarrow/.style={->, ultra thick, dashed}
]

\node[cloud] (audio) {Audio Input \\ (Amplitude $\in [-32768, 32767]$)};

\node[box, below=of audio] (digitencode) {Digit and Sign Encoding};
\node[below=0.1cm of digitencode, font=\footnotesize] {e.g.\ $+12345 \rightarrow \{0,1,2,3,4,5\}$};

\node[box, below=of digitencode] (spikeencode) {Spike Generation via LIF Neuron};
\node[below=0.1cm of spikeencode, font=\footnotesize] {NEST Simulation (60~ms Window)};

\node[box, below=of spikeencode] (mapping) {Modulo-16 Mapping and Key Assignment};
\node[below=0.1cm of mapping, font=\footnotesize] {Unique non-zero remainders per digit};

\node[box, below=of mapping] (embedding) {Image Embedding \\ (2 LSBs per RGBA Channel)};
\node[below=0.1cm of embedding, font=\footnotesize] {Stego-image Generation};

\node[hw, below=of embedding] (fpga) {FPGA (PYNQ-Z2) \\ On-Chip Encryption and Decryption};

\node[cloud, below=of fpga] (recaudio) {Recovered Audio};

\node[cloud, right=2.8cm of embedding] (stegoimg) {Stego-Image};

\draw[arrow] (audio) -- (digitencode);
\draw[arrow] (digitencode) -- (spikeencode);
\draw[arrow] (spikeencode) -- (mapping);
\draw[arrow] (mapping) -- (embedding);
\draw[arrow] (embedding) -- (fpga);
\draw[arrow] (fpga) -- (recaudio);
\draw[arrow] (embedding.east) -- (stegoimg.west);

\draw[decorate,decoration={brace, mirror, amplitude=6pt}, thick]
      ([xshift=-0.8cm]spikeencode.north west) --
      ([xshift=-0.55cm]mapping.south west)
      node[midway, xshift=-1.5cm, align=center, font=\bfseries]
      {Spike-to-Key\\Encoding \\Stage};

\draw[decorate,decoration={brace, mirror, amplitude=6pt}, thick]
      ([xshift=-0.55cm]embedding.north west) --
      ([xshift=-0.55cm]fpga.south west)
      node[midway, xshift=-1.7cm, align=center, font=\bfseries]
      {Steganographic\\Embedding \\Stage};

\node[font=\small, below=0.12cm of fpga] {Hardware Realisation (Low-Power Edge Implementation)};

\end{tikzpicture}}
\caption{Flowchart of the proposed \textbf{SteganoSNN} framework (portrait layout). The pipeline proceeds from audio digitisation and spike-based encoding (LIF neuron, NEST) through modulo-16 mapping and key assignment, followed by steganographic embedding into RGBA images. A PYNQ-Z2 FPGA performs on-chip encryption/decryption. Brace annotations group the \emph{Spike-to-Key Encoding} and \emph{Steganographic Embedding} stages.}
\label{fig:steganosnn_architecture_portrait}
\end{figure*}

The proposed \textit{SteganoSNN} framework (see~\cref{fig:steganosnn_architecture_portrait}) introduces a neuromorphic-inspired steganographic method for concealing audio data within four-channel PNG images\footnote{The PNG images must include four \gls*{rgba} channels.}. Its encoding process emulates the temporal information processing of \glspl*{snn} and is structured into four sequential stages: (1) digitisation of each audio sample into signed decimal components, (2) conversion of these digits into spike trains generated by \gls*{lif} neurons simulated in NEST, (3) modulo-based encryption and key assignment from spike timing patterns, and (4) embedding of the encrypted payload into the cover image. During decoding, the embedded data are extracted, decrypted, and sequentially reconstructed to recover the original audio waveform. The complete source code is available at: \protect\url{https://github.com/Biswajitks1/SNN-Based-Image-and-Audio-Steganography-with-Encryption}

As illustrated in~\cref{fig:steganosnn_architecture_portrait}, data are embedded into the two \glspl*{lsb} of each \gls*{rgba} channel, providing a total payload capacity of 8~hidden~bits~per~pixel. To preserve perceptual fidelity and mitigate the visual artefacts commonly introduced by bit substitution, a lightweight dithering stage injects bounded pseudo-random noise before embedding. This process maintains smooth intensity transitions and suppresses block-edge artefacts, ensuring that the stego-images remain visually indistinguishable from their originals.

\subsection{Digit Conversion and Spike Pattern Mapping}
Each audio sample, with amplitude in the range $[-32{,}768, +32{,}767]$, is represented as a sign component followed by a sequence of decimal digits. The sign is encoded separately—$0$ for positive and $1$ for negative—followed by the digits of its magnitude. For instance, the sample $+12{,}345$ becomes the sequence ${0, 1, 2, 3, 4, 5}$.

To generate biologically inspired spike patterns, a single \gls*{lif} neuron model was simulated in the NEST environment under constant input current stimulation. The relationship between input current and the resulting number of spikes was first characterised using the procedure described in Algorithm~\ref{alg:lif-spike-characterisation}. The neuron model parameters used in this process are listed in Table~\ref{tab:neuron-params}. The neuron fires when the membrane potential exceeds the threshold, and the number of spikes recorded corresponds directly to the input current magnitude.

Each spike pattern generated corresponds to an integer digit ($1$–$9$). The input current level was systematically increased until ten unique spike responses were obtained, representing digits from $0$ to $9$. The algorithm also identifies the precise current intervals that cause a transition in firing count. The resulting spike trains were recorded and analysed to determine characteristic spike timings (in milliseconds) within a fixed 60~ms simulation window.

For encryption, each digit was assigned a distinct spike position within its pattern, chosen such that the modulo-16 remainder of the timestamp is both unique and non-zero. Table~\ref{tab:spike_pattern} summarises the used mapping, showing the spike positions, chosen timestamps, and their modulo-16 remainders, which form the cipher basis used in the subsequent steganographic encoding process.

\begin{algorithm}[]
\caption{Spike Count Characterisation of LIF Neuron using NEST Simulator}
\label{alg:lif-spike-characterisation}
\begin{algorithmic}[1]
\State Reset NEST kernel
\State Set verbosity level to 20
\State Initialise variables:
    \Statex \hspace{1em} $currents \gets [\ ]$, $spike\_counts \gets [\ ]$
    \Statex \hspace{1em} $current \gets 370.0$, $inc \gets 1.0$
    \Statex \hspace{1em} $current\_spikes\_values \gets [[0, 0]]$
    \Statex \hspace{1em} $actual\_number\_spikes \gets 0$
\State Define LIF neuron parameters

\While{$|current\_spikes\_values| < 10$}
    \State Reset NEST kernel
    \State Create LIF neuron with given parameters
    \State Create spike recorder and connect neuron to it
    \State Set neuron input current $\gets current$
    \State Simulate for $60.0$~ms
    \State Record number of spikes $num\_spikes$
    \State Append $(current, num\_spikes)$ to $(currents, spike\_counts)$
    \If{$num\_spikes = 0$ \textbf{and} $current > current\_spikes\_values[0][0]$}
        \State Update $current\_spikes\_values[0][0] \gets current$
    \ElsIf{$num\_spikes > actual\_number\_spikes$}
        \State Append $[current, num\_spikes]$ to $current\_spikes\_values$
        \State Update $actual\_number\_spikes \gets num\_spikes$
    \EndIf
    \State Increment $current \gets current + inc$
\EndWhile

\State Save $current\_spikes\_values$ to \texttt{current\_spikes\_values.npy}
\State Plot spike count vs.\ input current
\end{algorithmic}
\end{algorithm}

\begin{table}[]
\centering
\caption{LIF Neuron Parameters and Simulation Settings}
\label{tab:neuron-params}
\begin{tabular}{@{}lll@{}}
\toprule
\textbf{Parameter} & \textbf{Symbol / Variable} & \textbf{Value} \\ 
\midrule
Membrane capacitance & $C\_m$ & 250.0~pF \\
Membrane time constant & $\tau\_m$ & 10.0~ms \\
Refractory period & $t\_{ref}$ & 2.0~ms \\
Resting potential & $E\_L$ & 0.0~mV \\
Threshold potential & $V\_{th}$ & 20.0~mV \\
Reset potential & $V\_{reset}$ & 10.0~mV \\
Excitatory synaptic time constant & $\tau\_{syn}^{ex}$ & 0.5~ms \\
Inhibitory synaptic time constant & $\tau\_{syn}^{in}$ & 0.5~ms \\
Initial membrane potential & $V\_m$ & -70.0~mV \\
Calcium concentration & $Ca$ & 0.0 \\
\midrule
\textbf{Simulation time} & $T\_{sim}$ & 60.0~ms \\
\textbf{Current increment} & $\Delta I$ & 1.0~pA \\
\textbf{Initial current} & $I\_{start}$ & 370.0~pA \\
\textbf{Termination condition} & -- & 10 distinct spike levels reached \\
\bottomrule
\end{tabular}
\end{table}

\begin{table}[]
\caption{Spike patterns for each digit, selected timestamps, modulo-16 remainders, and key indices.}
\label{tab:spike_pattern}
\centering
\begin{tabular}{@{}llccc@{}}
\toprule
\textbf{Digit} & \textbf{Spike Positions} & \textbf{Chosen Timestamp} & \textbf{Remainder (mod 16)} & \textbf{KEY} \\ 
\midrule
0 & -- & 0 & 0 & -- \\
1 & 59 & 59 & 11 & 0 \\
2 & 39, 59 & 39 & 7 & 0 \\
3 & 31, 45, 59 & 45 & 13 & 1 \\
4 & 26, 37, 48, 60 & 37 & 5 & 1 \\
5 & 23, 32, 41, 50, 59 & 50 & 2 & 3 \\
6 & 20, 28, 36, 44, 52, 59 & 44 & 12 & 3 \\
7 & 18, 25, 32, 39, 46, 52, 59 & 52 & 4 & 5 \\
8 & 17, 23, 29, 35, 41, 47, 53, 59 & 41 & 9 & 4 \\
9 & 15, 21, 26, 32, 37, 43, 48, 54, 59 & 54 & 6 & 7 \\ 
\bottomrule
\end{tabular}
\end{table}

\subsection{Encryption Process}\label{subsec:encryption}
Each selected spike position is encrypted into a 4-bit value by computing its modulo-16 remainder. The modulo-16 choice aligns with the 4-bit representation (0–15) while fully covering the 61 discrete time steps in the simulation window. Smaller moduli such as 8 would cause remainder collisions, whereas larger ones (e.g. 32) would exceed 4 bits. Thus, modulo-16 provides an optimal trade-off between uniqueness and embedding efficiency. The remainder serves as the ciphertext for the corresponding digit. Digit~0, which produces no spikes, maps to a remainder of 0 and remains unencrypted. To enable deterministic decryption, a 4-bit \texttt{KEY} value indicates the ordinal index of the selected spike within the full pattern (counting from zero).

\begin{example}
For digit~6, the spike pattern is $\{20, 28, 36, 44, 52, 59\}$.  
The chosen timestamp $44$ yields $r = 44 \bmod 16 = 12$.  
Since $44$ is the fourth spike (index~3), the corresponding \texttt{KEY} is~3.  
Hence, the cipher–key pair is $(12, 3)$.
\end{example}

The resulting cipher–key pairs expand the key space to $16 \times N$ possibilities, where $N$ is the number of spikes per digit, enhancing resistance to brute-force attacks.

\subsection{Steganographic Embedding}\label{subsec:stegembed}
The embedding uses \gls*{rgba} images where each channel is 8 bits. By replacing the two \glspl*{lsb} of each channel, 8 bits per pixel are available for embedding. Before LSB substitution, a lightweight dithering process introduces bounded pseudo-random noise to reduce quantisation artefacts and preserve perceptual uniformity.

Each audio sample (one sign bit and five digits) produces six encrypted symbols. Since each pixel stores 8 hidden bits, three pixels (24 bits) suffice to represent one audio channel sample, providing higher payload capacity than traditional \gls*{lsb} methods.

\begin{example}
The sample $+12{,}345$ yields digits $\{0, 1, 2, 3, 4, 5\}$, which map to encrypted values $\{0, 11, 7, 13, 5, 2\}$ (from Table~\ref{tab:spike_pattern}).  
Their 4-bit binary representations form:
\[
0000\,1011\,0111\,1101\,0101\,0010.
\]
Three pixels (24 bits total) embed this sequence across RGBA channels.  
Table~\ref{tab:example_embedding} shows one example after dithering and bit replacement.
\end{example}

\begin{table}[]
\centering
\caption{Example of embedding 8 audio bits into one \gls*{rgba} pixel with dithering noise.}
\label{tab:example_embedding}
\begin{tabular}{@{}lcccc@{}}
\toprule
\textbf{Channel} & \textbf{Original (Dec)} & \textbf{After Noise} & \textbf{LSBs Replaced} & \textbf{Modified (Dec + Binary)} \\ 
\midrule
R & 150 & 151 ($10010111$) & 00 & 148 ($10010100$) \\
G & 200 & 201 ($11001001$) & 00 & 200 ($11001000$) \\
B & 75  & 76  ($01001100$) & 10 & 74  ($01001010$) \\
A & 253 & 254 ($11111110$) & 11 & 255 ($11111111$) \\ 
\bottomrule
\end{tabular}
\end{table}

\subsection{Decryption Process}\label{subsec:decryption}
During retrieval, the embedded remainders are extracted from the LSBs and mapped back to their corresponding spike positions. For each remainder $r$, potential spike positions are computed as:
\[
r, \; r+16, \; r+32, \; r+48,
\]
bounded within the 0–60 window. The decoder identifies the digit whose spike pattern contains one of these positions. If multiple matches exist, the pre-shared \texttt{KEY} resolves ambiguity by specifying the index of the correct spike. Once all digits (and the sign) are recovered, the original 16-bit sample is reconstructed.

\begin{example}
Extracted LSBs $\{00, 00, 10, 11\}$ correspond to remainders $\{0, 11\}$.  
Remainder 0 maps to digit 0.  
For remainder 11, candidate spike positions $\{11, 27, 43, 59\}$ are compared to the reference patterns.  
Only digit 1 satisfies both the candidate set and the \texttt{KEY} constraint, confirming the correct value.  
Repeating this process yields all digits of the audio sample.
\end{example}

\subsection{Software Implementation}
The framework\footnote{The complete source code is available at: \protect\url{https://github.com/Biswajitks1/SNN-Based-Image-and-Audio-Steganography-with-Encryption}} was developed in Python with NEST for \gls*{lif} neuron simulation. It comprises modular scripts for pattern generation, key mapping, image preprocessing, embedding, and decryption.  
\texttt{Discretelevel.py} determines current thresholds; \texttt{SNN\_patterns.py} produces binary spike matrices; and \texttt{Key\_and\_Map.py} generates JSON-based digit–key maps.  
\texttt{Image\_analysis.py} ensures RGBA compliance and pre-masks LSBs; \texttt{encrypt.py} embeds ciphertexts, and \texttt{decrypt.py} reconstructs the audio waveform.

\begin{algorithm}[]
\caption{Software Workflow of \textit{SteganoSNN}}
\label{alg:software_workflow}
\begin{algorithmic}[1]
\Require Audio (\texttt{.wav}), cover image (\texttt{.png})
\Ensure Stego-image, losslessly recovered audio

\State \textbf{Init}:\; Init NEST;\; set \gls*{lif} params;\; define window \(t=0{:}60\).
\State \textbf{Currents}:\; Find discrete input currents (\texttt{Discretelevel.py}).
\State \textbf{Patterns}:\; Gen spike patterns for digits \(0\!-\!9\) (\texttt{SNN\_patterns.py}).
\State \textbf{Map/Key}:\; Compute mod-16 remainders;\; assign 4-bit \texttt{KEY} (index) (\texttt{Key\_and\_Map.py}); export JSON.
\State \textbf{Preprocess Img}:\; Ensure RGBA;\; dither (bounded);\; mask 2 LSBs (\texttt{Image\_analysis.py}).
\State \textbf{Embed}:\; Digitise samples (sign+digits)\(\rightarrow\)ciphertext;\; write to RGBA LSBs (\texttt{encrypt.py}); save stego.
\State \textbf{Extract}:\; Read RGBA LSBs;\; recover remainders (\texttt{decrypt.py}).
\State \textbf{Decode}:\; Use \texttt{KEY}+map\(\rightarrow\)digits;\; reassemble 16-bit samples; save audio.
\Return Stego-image,\; reconstructed audio
\end{algorithmic}
\end{algorithm}

\subsection{FPGA Implementation}
The framework was deployed on a PYNQ-Z2 FPGA board\footnote{The details about the board: \protect\url{https://pynq.readthedocs.io/en/latest/getting_started/pynq_z2_setup.html}}, which integrates a dual-core ARM Cortex-A9 (PS) with Xilinx Artix-7 programmable logic (PL). The PS manages data buffers, coordinates spike generation via NEST, and streams data to custom PL IP cores through AXI-DMA channels. The PL implements hardware accelerators for encryption and decryption, connected via AXI-Stream interfaces. The final co-design is packaged as a PYNQ overlay, enabling Python-level control of hardware execution.

To ensure consistency between software and hardware, NEST was compiled natively on the ARM subsystem. The design supports full-HD (1920×1080) RGBA images as carrier media and processes 16-bit audio samples ($-32{,}768$ to $+32{,}767$) represented as a sign plus five digits.

The PL comprises two primary IP cores: an \textit{Encryptor} and a \textit{Decryptor}.  
The Encryptor integrates a Digit Extractor implementing the Binary-to-BCD (Double-Dabble) algorithm, a lightweight pseudo-noise generator (0–2 cyclic offset), and Verilog modules implementing the encryption logic and embedding pipeline described earlier.  
The Decryptor performs symmetric extraction, LSB decoding, and spike-based digit reconstruction.

During execution, the PS first generates spike patterns (\texttt{SNN\_patterns.ipynb}), then loads the Encryptor bitstream (\texttt{final\_v1\_45\_encrypt.bit}) and prepares an interleaved data stream—three pixels per audio sample—transferred via AXI-DMA. The Encryptor outputs modified pixel data and a stored \texttt{KEY} for later decoding. The Decryptor bitstream (\texttt{final\_v1\_8\_decrypt.bit}) is then loaded, which reconstructs the audio stream from the encrypted image and the saved key values. All inter-module data transfers are handled as 32-bit unsigned streams for optimal DMA throughput.

\begin{algorithm}[]
\caption{Execution Flow on PYNQ-Z2 during Encryption and Decryption}
\label{alg:fpga_flow}
\begin{algorithmic}[1]
\Require Precomputed spike patterns; audio samples; cover image
\Ensure Encrypted stego-image and reconstructed audio

\State \textbf{Initialisation:}
\State Generate spike patterns via NEST on PS; store in DDR memory.

\State \textbf{Encryption Phase:}
\State Load \texttt{Encryptor} bitstream; allocate AXI-DMA buffers.
\State Interleave data (3 pixels : 1 audio sample) and stream to PL.
\State Execute encryption; output modified pixels and \texttt{KEY}.
\State Store encrypted image and \texttt{KEY} in DDR.

\State \textbf{Decryption Phase:}
\State Load \texttt{Decryptor} bitstream; reuse stored spike patterns.
\State Stream encrypted image and \texttt{KEY} to PL via DMA.
\State Execute decryption; reconstruct and output original audio.

\Return Stego-image and recovered audio waveform
\end{algorithmic}
\end{algorithm}

\section{Results and Discussion}\label{sec:results}

This section presents the performance evaluation of the proposed \textit{SteganoSNN} framework across key metrics, including encoding capacity, visual fidelity, steganalysis resistance, and hardware resource utilisation. Both hardware and software implementations produce identical functional outputs; the analysis and quantitative metrics presented in this section are derived from the software results.

\subsection{Datasets and Evaluation Environment}
The evaluation of the proposed SteganoSNN framework was performed using the DIV2K 2017 dataset~\cite{DIV2K2017} for cover images and standard 16-bit PCM .wav audio files as payloads. Each cover image was converted to \gls*{rgba} format before embedding. While the software framework can process images of arbitrary resolution, the hardware implementation on the PYNQ-Z2 platform supports image resolutions up to 1920~$\times$~1080. For quantitative analysis, \gls*{psnr} and \gls*{ssim} were computed between cover and stego-images to evaluate perceptual quality. These results are compared with SteganoGAN~\cite{zhang2019steganogan}. The Aletheia toolchain~\cite{Aletheia} was employed for steganalysis, using \gls*{spa}, Triples Analysis, and \gls*{ws}.  
    
\subsection{Steganographic Capacity}
The proposed approach achieved a high embedding capacity of 8 hidden BPP by modifying the two \glspl*{lsb} in each of the four \gls*{rgba} channels. This capacity exceeded that of conventional \gls*{lsb}-based steganography methods~\cite{neeta2006implementation}, The highest SteganoGAN~\cite{steganogan} could achieve is 5 bits per pixel. Although the experiments primarily focused on embedding audio data into images, the proposed encoding framework was data-agnostic. Because any digital information can be represented as numerical sequences, the same methodology can be extended to hide text, sensor readings, or other binary data within images, making the \textit{SteganoSNN} framework suitable for a wide range of multimodal data-hiding applications. In addition to image-based metrics, reconstructed audio samples were compared with their originals using waveform overlay and correlation analysis. The recovered audio showed 100\% bitwise accuracy, confirming lossless decoding. 

The audio payloads were 16-bit PCM \texttt{.wav} files sampled at 48~kHz with two channels (stereo). 
At an embedding rate of 8~hidden~bits per pixel (1~byte/pixel), a 1920$\times$1080 cover image provides approximately 2.07~MB of payload capacity. Given the stereo 48~kHz sampling rate, this corresponds to roughly 11~seconds of audio data that can be embedded within a single full-HD image. The sound~\cref{fig:audio_recovery} used here is of 6~seconds.
    
\subsection{Visual Fidelity and Artifact Suppression}
To maintain visual quality, a lightweight dithering stage was applied before \gls*{lsb} substitution. The controlled pseudo-random noise introduced during dithering preserved smooth gradients and suppressed visible artifacts that typically result from direct \gls*{lsb} modification. During initial design and hardware implementation, the dithering range was fixed to 0--2. Subsequent software experiments explored broader ranges and revealed that values up to 0--3 also yielded acceptable results. This parameter can be further studied, as different noise patterns or distributions may influence perceptual quality. Visual fidelity was quantitatively assessed using \gls*{psnr} and \gls*{ssim} on the DIV2K dataset~\cite{DIV2K2017}. The results confirmed minimal perceptual degradation between cover and stego-images, with \gls*{psnr} consistently exceeding 41 dB and \gls*{ssim} values above 0.97.

\subsection{Comparison with SteganoGAN}
The performance of the proposed method was evaluated using the Aletheia toolchain on the DIV2K 2017~\cite{DIV2K2017} dataset, and the results are summarised in Table~\ref{tab:aletheia_results}. Across all subsets, the proposed method maintains high perceptual quality and low distortion levels. The \gls*{psnr}\textsubscript{RGB} values range from 40.42 to 41.35~dB, while the corresponding \gls*{ssim}\textsubscript{RGB} values remain consistently high between 0.9693 and 0.9801, indicating minimal visual degradation. Similarly, the \gls*{psnr}\textsubscript{RGBA} values lie between 40.48 and 41.10~dB, with \gls*{ssim}\textsubscript{RGBA} scores above 0.97 across all subsets, confirming that the inclusion of the alpha channel does not compromise image quality.

\begin{table}[]
\centering
\caption{Performance comparison between SteganoGAN (dense model, $D{=}1\text{--}6$) and the proposed method on DIV2K. 
\emph{RS-BPP} is payload (hidden bits per pixel). For each metric/column, \best{best} values are bold green and \worst{worst} values are red underlined. 
SteganoGAN results are from~\cite{steganogan} and our proposed methods were benchmarked against DIV2K~2017~\cite{DIV2K2017}.}
\label{tab:comparison_grouped}

\begin{tabularx}{\linewidth}{@{}P{3cm} L c c c @{}}
\toprule
\textbf{Method} & \shortstack[l]{$D$\,/\\ Subset} & \textbf{RS-BPP} &
\textbf{PSNR$_{\text{RGB}}$} & \textbf{SSIM$_{\text{RGB}}$} \\
\midrule
\multirow{6}{*}{SteganoGAN (Dense)}
& $D{=}1$ & \worst{0.99} & 41.60 & 0.95 \\
& $D{=}2$ & 1.96 & 39.62 & 0.92 \\
& $D{=}3$ & 2.63 & \worst{36.52} & \worst{0.85} \\
& $D{=}4$ & 2.53 & 37.49 & 0.88 \\
& $D{=}5$ & 2.50 & 38.65 & 0.90\\
& $D{=}6$ & 2.44 & 38.84 & 0.90 \\
\midrule
\multirow{12}{*}{Ours}
& DIV2K\_train\_LR\_bicubic\_X2  & \best{8} & 41.01 & 0.9734 \\
& DIV2K\_train\_LR\_bicubic\_X3  & \best{8} & 41.13 & 0.9765 \\
& DIV2K\_train\_LR\_bicubic\_X4  & \best{8} & 41.34 & 0.9789 \\
& DIV2K\_train\_LR\_unknown\_X2  & \best{8} & 41.01 & 0.9693 \\
& DIV2K\_train\_LR\_unknown\_X3  & \best{8} & 41.12 & 0.9743 \\
& DIV2K\_train\_LR\_unknown\_X4  & \best{8} & 41.33 & 0.9748 \\
&  DIV2K\_valid\_LR\_bicubic\_X2 & \best{8} & 41.03 & 0.9745 \\
& DIV2K\_valid\_LR\_bicubic\_X3  & \best{8} & 41.15 & 0.9776 \\
& DIV2K\_valid\_LR\_bicubic\_X4  & \best{8} & \best{41.35} & \best{0.9801} \\
& DIV2K\_valid\_LR\_unknown\_X2   & \best{8} & 41.02 & 0.9705 \\
& DIV2K\_valid\_LR\_unknown\_X3   & \best{8} & 41.14 & 0.9756 \\
& DIV2K\_valid\_LR\_unknown\_X4   & \best{8} & 41.34 & 0.9762 \\
\bottomrule
\end{tabularx}
\end{table}
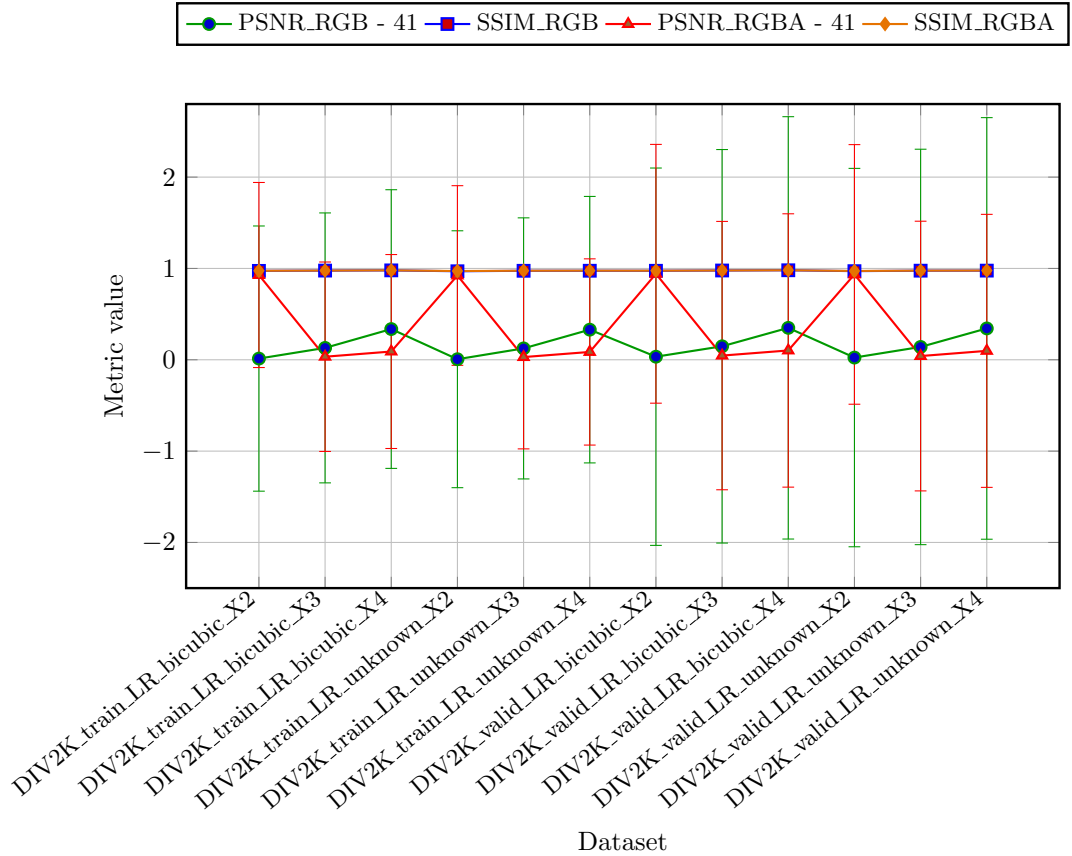
\begin{figure}[]
\centering
\begin{tikzpicture}
\begin{axis}[
    width=\textwidth,
    height=8cm,
    xlabel={Dataset},
    ylabel={Metric value},
    ymin=-2.5, ymax=2.8, 
    ymajorgrids,
    grid=major,
    legend style={at={(0.5,1.12)}, anchor=south, legend columns=-1, font=\small},
    error bars/y dir=both,
    error bars/y explicit,
    symbolic x coords={
      DIV2K_train_LR_bicubic_X2,
      DIV2K_train_LR_bicubic_X3,
      DIV2K_train_LR_bicubic_X4,
      DIV2K_train_LR_unknown_X2,
      DIV2K_train_LR_unknown_X3,
      DIV2K_train_LR_unknown_X4,
      DIV2K_valid_LR_bicubic_X2,
      DIV2K_valid_LR_bicubic_X3,
      DIV2K_valid_LR_bicubic_X4,
      DIV2K_valid_LR_unknown_X2,
      DIV2K_valid_LR_unknown_X3,
      DIV2K_valid_LR_unknown_X4
    },
    xtick=data,
    xticklabels={
      DIV2K\_train\_LR\_bicubic\_X2,
      DIV2K\_train\_LR\_bicubic\_X3,
      DIV2K\_train\_LR\_bicubic\_X4,
      DIV2K\_train\_LR\_unknown\_X2,
      DIV2K\_train\_LR\_unknown\_X3,
      DIV2K\_train\_LR\_unknown\_X4,
      DIV2K\_valid\_LR\_bicubic\_X2,
      DIV2K\_valid\_LR\_bicubic\_X3,
      DIV2K\_valid\_LR\_bicubic\_X4,
      DIV2K\_valid\_LR\_unknown\_X2,
      DIV2K\_valid\_LR\_unknown\_X3,
      DIV2K\_valid\_LR\_unknown\_X4
    },
    xticklabel style={rotate=40, anchor=east, font=\scriptsize},
    line width=0.9pt,
    mark size=2.2pt,
]

\addplot+[green!60!black, mark=*, thick, error bars/.cd, y dir=both, y explicit]
coordinates {
    (DIV2K_train_LR_bicubic_X2,0.0132)+-(0.5839,1.4518)
    (DIV2K_train_LR_bicubic_X3,0.1301)+-(0.5845,1.4778)
    (DIV2K_train_LR_bicubic_X4,0.3364)+-(0.5858,1.5255)
    (DIV2K_train_LR_unknown_X2,0.0060)+-(0.5825,1.4064)
    (DIV2K_train_LR_unknown_X3,0.1247)+-(0.5893,1.4293)
    (DIV2K_train_LR_unknown_X4,0.3297)+-(0.5984,1.4588)
    (DIV2K_valid_LR_bicubic_X2,0.0337)+-(0.2527,2.0658)
    (DIV2K_valid_LR_bicubic_X3,0.1476)+-(0.2338,2.1542)
    (DIV2K_valid_LR_bicubic_X4,0.3495)+-(0.2993,2.3121)
    (DIV2K_valid_LR_unknown_X2,0.0242)+-(0.2325,2.0715)
    (DIV2K_valid_LR_unknown_X3,0.1406)+-(0.2315,2.1648)
    (DIV2K_valid_LR_unknown_X4,0.3433)+-(0.2682,2.3080)
};
\addlegendentry{PSNR\_RGB - 41}

\addplot+[blue, mark=square*, thick, error bars/.cd, y dir=both, y explicit]
coordinates {
    (DIV2K_train_LR_bicubic_X2,0.9734)+-(0.0918,0.0234)
    (DIV2K_train_LR_bicubic_X3,0.9765)+-(0.0854,0.0206)
    (DIV2K_train_LR_bicubic_X4,0.9789)+-(0.0768,0.0194)
    (DIV2K_train_LR_unknown_X2,0.9693)+-(0.0911,0.0242)
    (DIV2K_train_LR_unknown_X3,0.9743)+-(0.0848,0.0228)
    (DIV2K_train_LR_unknown_X4,0.9748)+-(0.0746,0.0225)
    (DIV2K_valid_LR_bicubic_X2,0.9745)+-(0.0510,0.0213)
    (DIV2K_valid_LR_bicubic_X3,0.9776)+-(0.0428,0.0201)
    (DIV2K_valid_LR_bicubic_X4,0.9801)+-(0.0343,0.0182)
    (DIV2K_valid_LR_unknown_X2,0.9705)+-(0.0525,0.0238)
    (DIV2K_valid_LR_unknown_X3,0.9756)+-(0.0451,0.0215)
    (DIV2K_valid_LR_unknown_X4,0.9762)+-(0.0375,0.0181)
};
\addlegendentry{SSIM\_RGB}

\addplot+[red, mark=triangle*, thick, error bars/.cd, y dir=both, y explicit]
coordinates {
    (DIV2K_train_LR_bicubic_X2,0.9281)+-(0.4368,1.0132)
    (DIV2K_train_LR_bicubic_X3,0.0335)+-(0.4347,1.0371)
    (DIV2K_train_LR_bicubic_X4,0.0904)+-(0.4250,1.0617)
    (DIV2K_train_LR_unknown_X2,0.9228)+-(0.4359,0.9834)
    (DIV2K_train_LR_unknown_X3,0.0296)+-(0.4386,1.0051)
    (DIV2K_train_LR_unknown_X4,0.0857)+-(0.4344,1.0195)
    (DIV2K_valid_LR_bicubic_X2,0.9415)+-(0.1853,1.4167)
    (DIV2K_valid_LR_bicubic_X3,0.0459)+-(0.1704,1.4691)
    (DIV2K_valid_LR_bicubic_X4,0.1020)+-(0.2183,1.4964)
    (DIV2K_valid_LR_unknown_X2,0.9346)+-(0.1704,1.4212)
    (DIV2K_valid_LR_unknown_X3,0.0408)+-(0.1689,1.4764)
    (DIV2K_valid_LR_unknown_X4,0.0977)+-(0.1943,1.4946)
};
\addlegendentry{PSNR\_RGBA - 41}

\addplot+[orange!90!black, mark=diamond*, thick, error bars/.cd, y dir=both, y explicit]
coordinates {
    (DIV2K_train_LR_bicubic_X2,0.9747)+-(0.0688,0.0176)
    (DIV2K_train_LR_bicubic_X3,0.9768)+-(0.0641,0.0155)
    (DIV2K_train_LR_bicubic_X4,0.9791)+-(0.0577,0.0145)
    (DIV2K_train_LR_unknown_X2,0.9716)+-(0.0682,0.0182)
    (DIV2K_train_LR_unknown_X3,0.9752)+-(0.0637,0.0166)
    (DIV2K_train_LR_unknown_X4,0.9760)+-(0.0561,0.0169)
    (DIV2K_valid_LR_bicubic_X2,0.9755)+-(0.0382,0.0160)
    (DIV2K_valid_LR_bicubic_X3,0.9776)+-(0.0320,0.0151)
    (DIV2K_valid_LR_bicubic_X4,0.9800)+-(0.0257,0.0136)
    (DIV2K_valid_LR_unknown_X2,0.9725)+-(0.0394,0.0179)
    (DIV2K_valid_LR_unknown_X3,0.9761)+-(0.0337,0.0161)
    (DIV2K_valid_LR_unknown_X4,0.9770)+-(0.0279,0.0136)
};
\addlegendentry{SSIM\_RGBA}

\end{axis}
\end{tikzpicture}

\caption{
Error bar plot showing minimum, mean, and maximum values for PSNR\_RGB, SSIM\_RGB, PSNR\_RGBA, and SSIM\_RGBA metrics across DIV2K 2017 dataset. PSNR values are shown **relative to 41** (PSNR - 41) to highlight small variations; SSIM values are plotted as original.
}
\label{fig:combined_psnr_ssim_zoomed}
\end{figure}

\subsection{Steganalysis Resistance}
Robustness against steganalysis was evaluated using the Aletheia toolchain~\cite{Aletheia}, considering three statistical detectors: \gls*{spa}, Triples Analysis, and \gls*{ws}. The results, summarised in Table~\ref{tab:aletheia_results}, demonstrate the stability and low detectability of the proposed embedding method across all DIV2K~2017 subsets.

Given the higher perceptual and statistical sensitivity of the Green channel, SPA\textsubscript{G}, Triples\textsubscript{G}, and WS\textsubscript{G} were analysed in detail. The SPA\textsubscript{G} metric ranged from 0.3362 (DIV2K\_train\_LR\_bicubic\_X4) to 0.7603 (DIV2K\_valid\_LR\_unknown\_X3), indicating a low probability of detection under second-order statistical attacks. Triples\textsubscript{G} values remained remarkably consistent at 0.7368 across all subsets, suggesting statistical uniformity and minimal embedding bias. WS\textsubscript{G} values varied between 1.1176 and 2.0519, with lower scores (e.g., DIV2K\_train\_LR\_bicubic\_X4) denoting smoother residual distributions and higher image naturalness. 

\begin{figure}[]
\centering
\resizebox{\textwidth}{!}{%
\begin{tikzpicture}
\begin{axis}[
  width=\textwidth,
  height=7.5cm,
  xlabel={Dataset categories},
  ylabel={Metric value},
  ymin=0, ymax=2.5,
  grid=both,
  legend style={at={(0.5,1.12)},anchor=north,legend columns=-1},
  symbolic x coords={
    trHR,trX2,trX3,trX4,
    trUX2,trUX3,trUX4,
    vaHR,vaX2,vaX3,vaX4,
    vaUX2,vaUX3,vaUX4
  },
  xticklabels={
    DIV2K\_train\_HR,
    DIV2K\_train\_LR\_bicubic\_X2,
    DIV2K\_train\_LR\_bicubic\_X3,
    DIV2K\_train\_LR\_bicubic\_X4,
    DIV2K\_train\_LR\_unknown\_X2,
    DIV2K\_train\_LR\_unknown\_X3,
    DIV2K\_train\_LR\_unknown\_X4,
    DIV2K\_valid\_HR,
    DIV2K\_valid\_LR\_bicubic\_X2,
    DIV2K\_valid\_LR\_bicubic\_X3,
    DIV2K\_valid\_LR\_bicubic\_X4,
    DIV2K\_valid\_LR\_unknown\_X2,
    DIV2K\_valid\_LR\_unknown\_X3,
    DIV2K\_valid\_LR\_unknown\_X4
  },
  x tick label style={rotate=35, anchor=east, font=\footnotesize},
  tick label style={font=\small},
  label style={font=\small},
  mark size=2.2pt,
  every axis plot/.append style={thick},
  error bars/y dir=both,
  error bars/y explicit,
]

\addplot+[mark=*, color=green!60!black]
coordinates {
  (trHR,0.35424625) +- (0.22624625,0.24975375)
  (trX2,0.519045)   +- (0.129045,0.068955)
  (trX3,0.74854875) +- (0.41854875,0.12045125)
  (trX4,0.33627625) +- (0.27727625,0.23072375)
  (trUX2,0.52394375) +- (0.09494375,0.06105625)
  (trUX3,0.75412)    +- (0.41112,0.11388)
  (trUX4,0.34809)    +- (0.25109,0.21091)
  (vaHR,0.33693)     +- (0.13593,0.25907)
  (vaX2,0.52267)     +- (0.07367,0.04433)
  (vaX3,0.75615)     +- (0.30015,0.08285)
  (vaX4,0.34827)     +- (0.20027,0.13873)
  (vaUX2,0.52693)    +- (0.05993,0.03507)
  (vaUX3,0.76032)    +- (0.31032,0.06868)
  (vaUX4,0.35908)    +- (0.16008,0.10592)
};
\addlegendentry{SPA (G)}

\addplot+[mark=square*, color=blue!70!black]
coordinates {
  (trHR,0.736842105263158) +- (0.0,0.0)
  (trX2,0.736842105263158) +- (0.0,0.0)
  (trX3,0.736842105263158) +- (0.0,0.0)
  (trX4,0.736842105263158) +- (0.0,0.0)
  (trUX2,0.736842105263158) +- (0.0,0.0)
  (trUX3,0.736842105263158) +- (0.0,0.0)
  (trUX4,0.736842105263158) +- (0.0,0.0)
  (vaHR,0.736842105263158) +- (0.0,0.0)
  (vaX2,0.736842105263158) +- (0.0,0.0)
  (vaX3,0.736842105263158) +- (0.0,0.0)
  (vaX4,0.736842105263158) +- (0.0,0.0)
  (vaUX2,0.736842105263158) +- (0.0,0.0)
  (vaUX3,0.736842105263158) +- (0.0,0.0)
  (vaUX4,0.736842105263158) +- (0.0,0.0)
};
\addlegendentry{Triples (G)}

\addplot+[mark=triangle*, color=red!70!black]
coordinates {
  (trHR,1.2090975) +- (0.7670975,0.6139025)
  (trX2,1.4643225) +- (0.3633225,0.2776775)
  (trX3,2.04348875) +- (1.12048875,0.37951125)
  (trX4,1.117625) +- (0.694625,0.489375)
  (trUX2,1.4849875) +- (0.3209875,0.2640125)
  (trUX3,2.05198) +- (1.07798,0.36902)
  (trUX4,1.16099375) +- (0.63999375,0.42200625)
  (vaHR,1.16521) +- (0.56721,0.57279)
  (vaX2,1.47166) +- (0.22066,0.16834)
  (vaX3,2.04093) +- (0.89793,0.21807)
  (vaX4,1.14711) +- (0.48211,0.36589)
  (vaUX2,1.49062) +- (0.20262,0.13638)
  (vaUX3,2.05014) +- (0.85414,0.16986)
  (vaUX4,1.18708) +- (0.35408,0.30092)
};
\addlegendentry{WS (G)}

\end{axis}
\end{tikzpicture}
e}
\caption{Combined error bar plot showing minimum, mean, and maximum values of the SPA, Triples, and WS metrics for the Green (G) channel across DIV2K 2017 dataset. 
Each curve connects mean values, while the vertical bars represent the range (minimum to maximum) observed per dataset. 
SPA (green) quantifies spatial pixel distortion, Triples (blue) measures triple-pixel correlation, and WS (red) captures wavelet-domain similarity. 
The Green channel is emphasised for its perceptual dominance in human vision.}
\label{fig:combined_g_errorbar}
\end{figure}
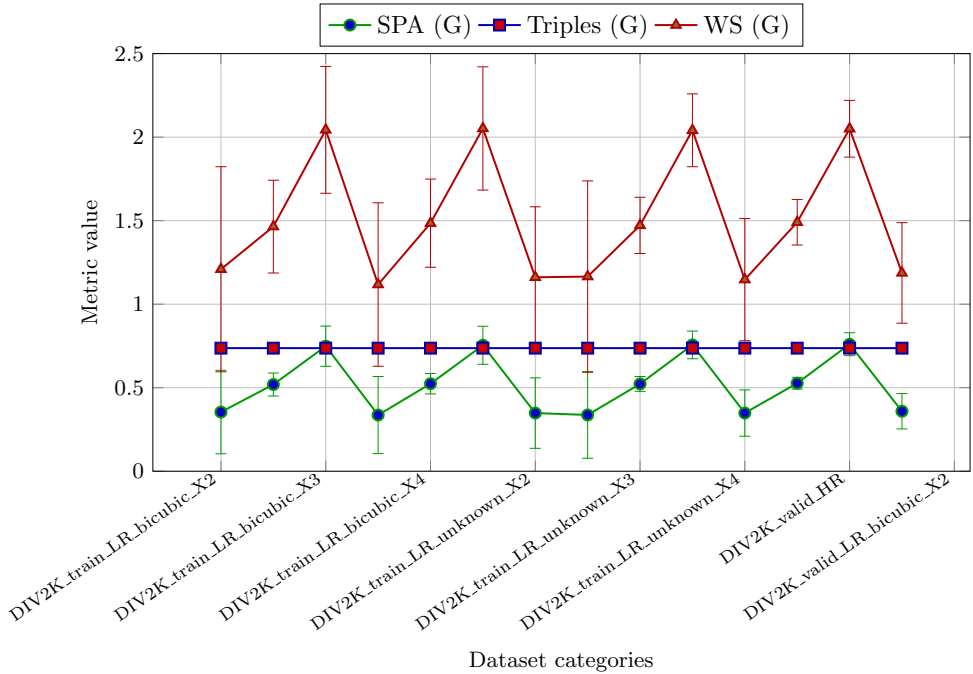

Overall, the results confirm that the proposed \textit{SteganoSNN} framework produces highly imperceptible modifications, achieving both pixel-wise and statistical invisibility across multiple scales and degradation types. The stability of the Triples and WS metrics further validates the resilience of the embedded data against modern steganalysis techniques. These findings are visually summarised in \cref{fig:combined_g_errorbar}, which presents the SPA, Triples, and WS distributions across all tested subsets.

\begin{landscape}

\begin{table}[]
\centering
\caption{Results obtained from the Aletheia toolchain on the DIV2K 2017. 
Best (lowest) results are highlighted in \cellcolor{green!20}{green} and worst (highest) in \cellcolor{red!20}{red}.}
\label{tab:aletheia_results}
\begin{tabular}{@{}lccccccccc@{}}
\toprule
\textbf{Folder} & SPA\_R & SPA\_G & SPA\_B & Triples\_R & Triples\_G & Triples\_B & WS\_R & WS\_G & WS\_B \\
\midrule
DIV2K\_train\_HR           
& \cellcolor{green!20}{0.1232} & 0.3542 & \cellcolor{green!20}{0.1189} 
& \cellcolor{green!20}{0.2149} & 0.7368 & \cellcolor{green!20}{0.2043} 
& \cellcolor{green!20}{0.2676} & 1.2091 & \cellcolor{green!20}{0.2111} \\

DIV2K\_train\_LR\_bicubic\_X2 
& 0.2031 & 0.5190 & 0.2282 
& 0.7368 & 0.7368 & 0.7368 
& 0.3858 & 1.4643 & 0.3777 \\

DIV2K\_train\_LR\_bicubic\_X3   
& 1.0105 & 0.7485 & 0.4853 
& 0.7301 & 0.7368 & 0.7359 
& 1.5008 & 2.0435 & 0.8077 \\

DIV2K\_train\_LR\_bicubic\_X4   
& 0.1488 & \cellcolor{green!20}{0.3362} & 0.1400 
& 0.6892 & 0.7368 & 0.7318 
& 0.3051 & \cellcolor{green!20}{1.1176} & 0.2150 \\

DIV2K\_train\_LR\_unknown\_X2   
& 0.2033 & 0.5239 & 0.2296 
& 0.7368 & 0.7368 & 0.7368 
& 0.39446 & 1.4850 & 0.3914 \\

DIV2K\_train\_LR\_unknown\_X3   
& 1.0245 & 0.7541 & \cellcolor{red!20}{0.4891} 
& 0.7345 & 0.7368 & 0.7368 
& \cellcolor{red!20}{1.5129} & \cellcolor{red!20}{2.0519} & 0.8166 \\

DIV2K\_train\_LR\_unknown\_X4 
& 0.1483 & 0.3480 & 0.1471 
& 0.7094 & 0.7368 & 0.7352 
& 0.3127 & 1.1609 & 0.2255 \\

DIV2K\_valid\_LR\_bicubic\_X2     
& 0.2013 & 0.5227 & 0.2258 
& 0.7368 & 0.7368 & 0.7368 
& 0.3856 & 1.4717 & 0.3788 \\

DIV2K\_valid\_LR\_bicubic\_X3     
& 1.0243 & 0.7561 & 0.4815 
& 0.7324 & 0.7368 & 0.7368 
& 1.4999 & 2.0409 & 0.8051 \\

DIV2K\_valid\_LR\_bicubic\_X4       
& 0.1528 & 0.3483 & 0.1400 
& 0.6969 & 0.7368 & \cellcolor{green!20}{0.7302} 
& 0.3079 & 1.1471 & 0.2175 \\

DIV2K\_valid\_LR\_unknown\_X2 
& 0.2008 & 0.5269 & 0.2268 
& 0.7368 & 0.7368 & 0.7368 
& 0.3949 & 1.4060 & 0.3940 \\

DIV2K\_valid\_LR\_unknown\_X3 
& \cellcolor{red!20}{1.0332} & \cellcolor{red!20}{0.7603} & 0.4863 
& 0.7368 & 0.7368 & \cellcolor{red!20}{0.7368} 
& 1.5122 & 2.0501 & \cellcolor{red!20}{0.8187} \\

DIV2K\_valid\_LR\_unknown\_X4 
& 0.1515 & 0.3591 & 0.1450 
& 0.7059 & 0.7368 & 0.7368 
& 0.3142 & 1.1871 & 0.2285 \\
\botrule
\end{tabular}
\end{table}

\end{landscape}

Because the Aletheia framework does not support analysis of the alpha (A) channel, quantitative results for that channel were not included; however, all four RGBA channels are used. These results indicate that the proposed embedding approach maintains low statistical detectability, even at a high payload of 8 \gls*{bpp}, demonstrating robustness against standard steganalysis methods.

\subsection{FPGA Resource Utilisation}
The implementation supports Full HD (1920×1080) PNG images. Resource utilisation of the custom \textit{Encryptor} and \textit{Decryptor} Intellectual Property (IP) cores is presented in Table~\ref{tab:ip_resource_utilization}. The Encryptor core occupies less than 10\% of available Look-Up Tables (LUTs), demonstrating excellent efficiency. The Decryptor utilises a higher LUT share (65.35\%) but requires minimal Flip-Flops (FFs), Block RAM (BRAM), and LUTRAM resources. The decryptor's LUT usage is elevated due to nested loops in the Verilog implementation.

High-speed data streaming between DDR memory and the IP cores is achieved using AXI DMA engines. The system further eliminates PC dependency by executing NEST simulations and \gls*{lif} neuron computations directly on the PYNQ-Z2’s ARM processor (Processing System, PS).
\begin{table}[]
\caption{Resource utilisation of Encryptor and Decryptor IP cores}
\label{tab:ip_resource_utilization}
  \begin{tabular}{@{}lrrrr|rrrr@{}}
 \toprule
  \multirow{2}{*}{\textbf{Resource Type}} & \multicolumn{3}{c}{\textbf{Encryptor}} & & \multicolumn{3}{c}{\textbf{Decryptor}} \\
 \cmidrule(lr){2-4} \cmidrule(lr){6-8}
    & \textbf{Utilisation} & \textbf{Available} & \textbf{(\%)} & & \textbf{Utilisation} & \textbf{Available} & \textbf{(\%)} \\
   \midrule
   LUT    & 5185  & 53200  & 9.75  & & 34768 & 53200 & 65.35 \\
   LUTRAM & 185   & 17400  & 1.06  & & 185   & 17400 & 1.06  \\
   FF     & 4266  & 106400 & 4.01  & & 3909  & 106400 & 3.67  \\
   BRAM   & 3     & 140    & 2.14  & & 3     & 140    & 2.14  \\
  BUFG   & 2     & 32     & 6.25  & & 1     & 32     & 3.13  \\
\bottomrule
\end{tabular}
\end{table}

\subsection{Stego-Image and Audio Recovery Results}
\begin{figure}[]
\centering
\begin{subfigure}{0.48\textwidth}
    \centering
    \includegraphics[width=\linewidth]{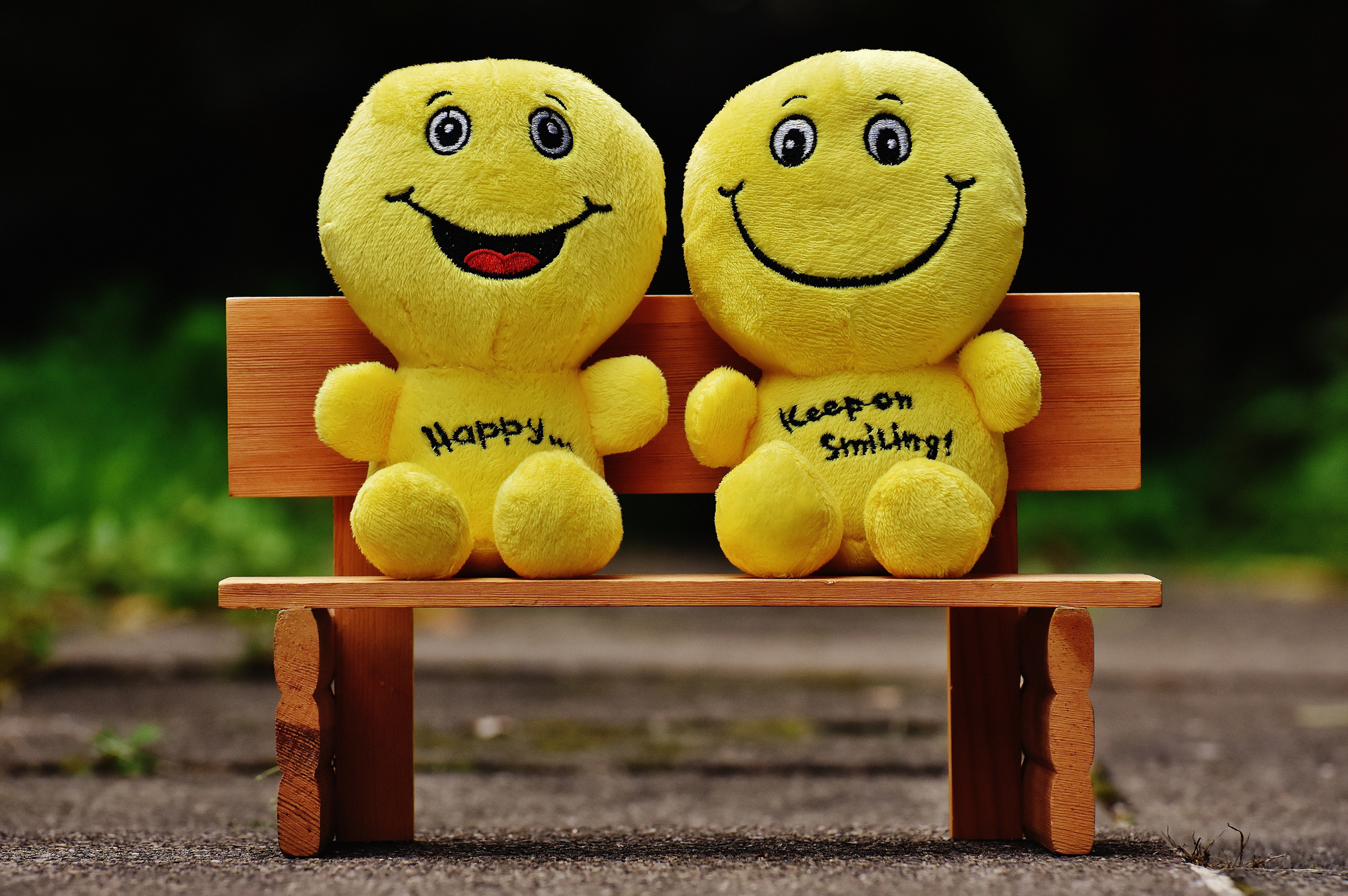}
    \caption{Original image.}
    \label{fig:original_image}
\end{subfigure}\hfill
\begin{subfigure}{0.48\textwidth}
    \centering
    \includegraphics[width=\linewidth]{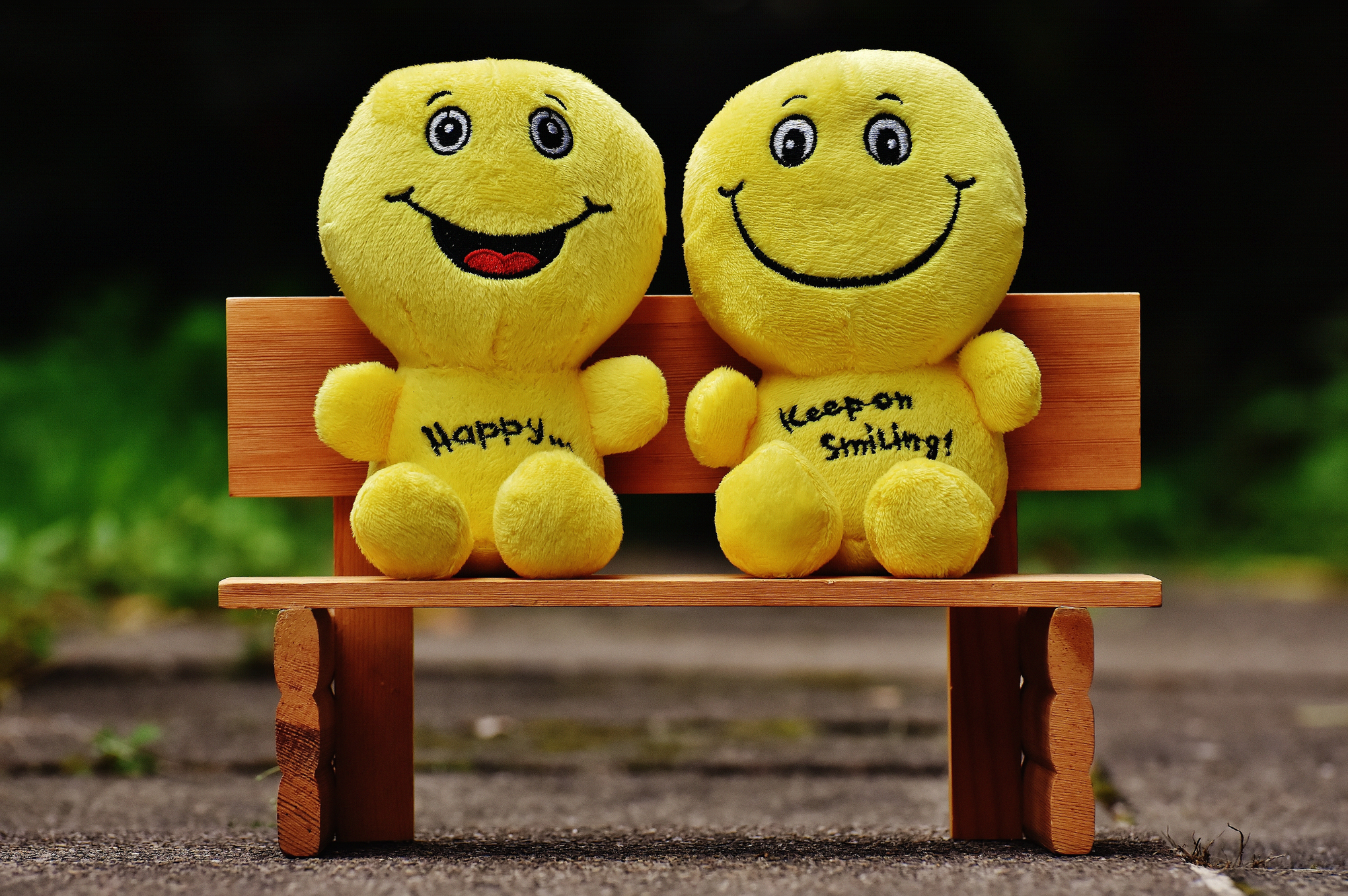}
    \caption{Stego-image after audio embedding.}
    \label{fig:stego_image}
\end{subfigure}
\caption{Comparison of (a) original and (b) stego-image. Visual differences are imperceptible, confirming high-fidelity embedding.}
\label{fig:img_comparison}
\end{figure}

To illustrate the performance of the proposed framework, we present a single high-resolution image from the \texttt{DIV2K\_train\_HR} dataset~\cite{DIV2K2017}. \cref{fig:original_image} shows the original cover image, and~\cref{fig:stego_image} shows the corresponding stego-image after embedding the audio. The visual quality remains nearly identical, confirming minimal perceptual distortion due to the embedding process.

\begin{figure}[]
\centering
\includegraphics[width=0.9\textwidth]{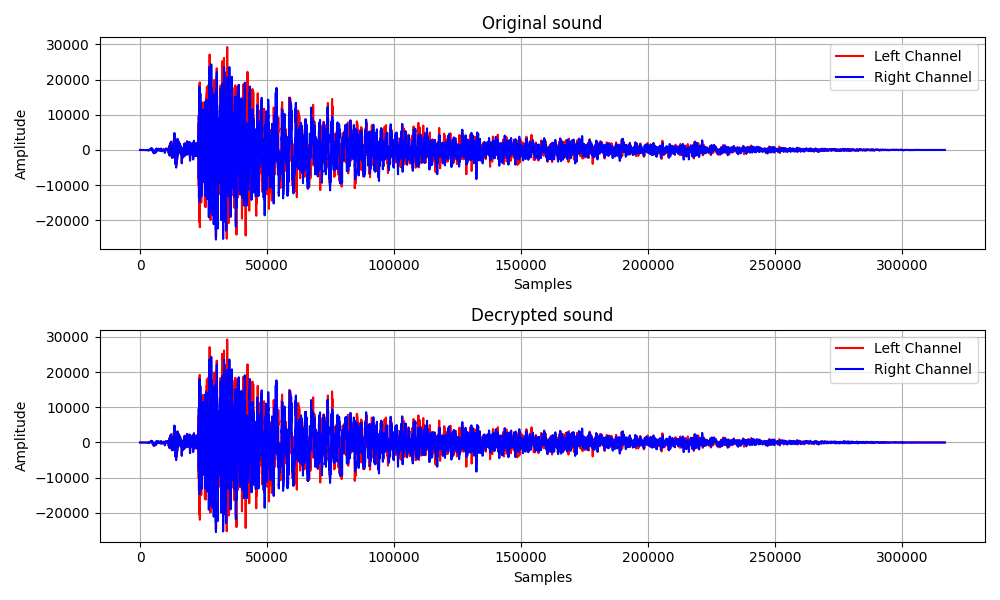}
\caption{Original and recovered audio waveforms showing lossless reconstruction.}
\label{fig:audio_recovery}
\end{figure}

\cref{fig:audio_recovery} shows the original and recovered audio waveforms plotted together. The matching waveforms demonstrate accurate recovery and validate the correctness of the audio embedding and decryption procedure.

The proposed \textit{SteganoSNN} framework effectively integrates \gls*{snn}-inspired spike-based encryption with high-capacity image steganography. Experimental results on the DIV2K~2017 dataset demonstrate a consistent embedding rate of 8~\gls*{bpp}, with visual fidelity maintained at a high level (\gls*{psnr} between 40.4~dB and 41.35~dB, \gls*{ssim}~$>$~0.97) across all training and validation subsets. The method exhibits strong resilience to statistical steganalysis, as indicated by low \gls*{spa} values (0.33–0.76), stable Triples scores (0.7368), and bounded \gls*{ws} metrics (1.12–2.05), confirming minimal statistical detectability. In addition, a functional \gls*{fpga}-based implementation validates the framework’s suitability for energy-efficient operation on embedded and edge devices. Collectively, these findings establish \textit{SteganoSNN} as a robust and neuromorphically inspired solution for secure, high-capacity multimedia data hiding in resource-constrained environments.

\section{Conclusions and Future Work}
This paper presented \textbf{SteganoSNN}, a biologically inspired steganographic framework that leverages \glspl*{snn} to enable secure, energy-efficient, and high-capacity data hiding. The proposed system encodes audio signals into spike-based representations using \gls*{lif} neurons simulated in the NEST environment. These spike trains are encrypted through a modulo-based mapping scheme and embedded into the least significant bits of \gls*{rgba} image channels using a controlled dithering mechanism to minimise visual distortion. The adoption of spiking neuron dynamics facilitates lightweight, low-power data-hiding architectures well suited for deployment on edge platforms such as \glspl*{fpga}.

Experimental evaluations on the DIV2K~2017 dataset confirmed that the proposed approach maintains high perceptual quality, achieving \gls*{psnr} values between 40.4 dB and 41.35 dB and \gls*{ssim} values consistently above 0.97 across all subsets. The resulting stego-images were visually indistinguishable from the originals, while the recovered audio signals exhibited perfect waveform reconstruction. Compared with SteganoGAN, \textit{SteganoSNN} achieved a substantially higher embedding capacity (8 \gls*{bpp}) with markedly lower computational overhead, underscoring its potential for secure and real-time multimedia communication in resource-constrained environments.

Beyond its technical merits, this work introduces a new research direction in neuromorphic steganography, demonstrating that spike-based encoding can simultaneously support secure communication and hardware efficiency. The modular implementation, verified both in software and on a PYNQ-Z2 \gls*{fpga}, illustrates the feasibility of neuromorphic architectures for multimedia encryption tasks.

Future work will focus on three main avenues. First, expanding the framework to support \gls*{snn}-based image-to-image or multimodal embedding, enabling secure transmission of video and sensor streams. Second, integrating on-chip learning capabilities through local plasticity mechanisms (e.g., STDP) to enhance adaptability against adversarial steganalysis attacks. Third, exploring spike-timing-based encryption schemes and reservoir computing architectures for improving the robustness and information density of spike-coded data. These extensions will strengthen the practical relevance of SteganoSNN in secure edge computing, medical telemetry, and autonomous system communications.

\bibliography{myBib}

@InProceedings{Plagwitz2024,
  title={SNN vs. CNN implementations on FPGAs: an empirical evaluation},
  author={Plagwitz, Patrick and Hannig, Frank and Teich, J{\"u}rgen and Keszocze, Oliver},
  booktitle={International Symposium on Applied Reconfigurable Computing},
  pages={3--18},
  year={2024},
  organization={Springer}
}

@ARTICLE{9787485,
  title={Spiking neural networks: A survey},
  author={Nunes, Jo{\~a}o D and Carvalho, Marcelo and Carneiro, Diogo and Cardoso, Jaime S},
  journal={IEEE access},
  volume={10},
  pages={60738--60764},
  year={2022},
  publisher={IEEE}
}

@inproceedings{Stewart_2022,
   title={Encoding event-based data with a hybrid snn guided variational auto-encoder in neuromorphic hardware},
  author={Stewart, Kenneth and Danielescu, Andreea and Shea, Timothy and Neftci, Emre},
  booktitle={Proceedings of the 2022 Annual Neuro-Inspired Computational Elements Conference},
  pages={88--97},
  year={2022}
}

@misc{lv2024efficient,
  title={Efficient and effective time-series forecasting with spiking neural networks},
  author={Lv, Changze and Wang, Yansen and Han, Dongqi and Zheng, Xiaoqing and Huang, Xuanjing and Li, Dongsheng},
  journal={arXiv preprint arXiv:2402.01533},
  year={2024}
}

@misc{venzke2020artificial,
  title={Artificial neural networks for sensor data classification on small embedded systems},
  author={Venzke, Marcus and Klisch, Daniel and Kubik, Philipp and Ali, Asad and Missier, Jesper Dell and Turau, Volker},
  journal={arXiv preprint arXiv:2012.08403},
  year={2020}
}

@misc{malcolm2023comprehensive,
  title={A comprehensive review of spiking neural networks: Interpretation, optimization, efficiency, and best practices},
  author={Malcolm, Kai and Casco-Rodriguez, Josue},
  journal={arXiv preprint arXiv:2303.10780},
  year={2023}
}

@inproceedings{shi2024towards,
  title={Towards energy efficient spiking neural networks: An unstructured pruning framework},
  author={Shi, Xinyu and Ding, Jianhao and Hao, Zecheng and Yu, Zhaofei},
  booktitle={The Twelfth International Conference on Learning Representations},
  year={2024}
}

@ARTICLE{1333071,
  title={Which model to use for cortical spiking neurons?},
  author={Izhikevich, Eugene M},
  journal={IEEE transactions on neural networks},
  volume={15},
  number={5},
  pages={1063--1070},
  year={2004},
  publisher={Ieee}
}

@article{Hodgkin1952,
title={A quantitative description of membrane current and its application to conduction and excitation in nerve},
  author={Hodgkin, Alan L and Huxley, Andrew F},
  journal={The Journal of physiology},
  volume={117},
  number={4},
  pages={500},
  year={1952}
}

@article{burkitt,
  title={A review of the integrate-and-fire neuron model: I. Homogeneous synaptic input},
  author={Burkitt, Anthony N},
  journal={Biological cybernetics},
  volume={95},
  number={1},
  pages={1--19},
  year={2006},
  publisher={Springer}
}

@inbook{Gerstner_Kistler_2002, 
  title={Spiking neuron models: Single neurons, populations, plasticity},
  author={Gerstner, Wulfram and Kistler, Werner M},
  year={2002},
  publisher={Cambridge university press}
}

@misc{lu2022linear,
  title={Linear leaky-integrate-and-fire neuron model based spiking neural networks and its mapping relationship to deep neural networks},
  author={Lu, Sijia and Xu, Feng},
  journal={Frontiers in neuroscience},
  volume={16},
  pages={857513},
  year={2022},
  publisher={Frontiers Media SA}
}

@ARTICLE{1257420,
  title={Simple model of spiking neurons},
  author={Izhikevich, Eugene M},
  journal={IEEE Transactions on neural networks},
  volume={14},
  number={6},
  pages={1569--1572},
  year={2003},
  publisher={IEEE}
}

@ARTICLE{Gewaltig:2007,
  author       = {Terhorst, Dennis and
                  Zajzon, Barna and
                  Vogelsang, Jan and
                  Korcsak-Gorzo, Agnes and
                  Lober, Melissa and
                  Espinoza Valverde, Jesus Andres and
                  Rechl, Maximilian and
                  Jiang, Han-Jia and
                  Linssen, Charl and
                  Kunkel, Susanne and
                  Graber, Steffen and
                  Müller, Eric and
                  Trensch, Guido and
                  Skaar, Jan-Eirik Welle and
                  Mitchell, Jessica and
                  Spreizer, Sebastian and
                  Benelhedi, Ayssar and
                  Serenko, Alexey and
                  Lee, Anthony Y and
                  Vorobev, Viktor and
                  Brunelli, Elide and
                  Madhav, Mahesh and
                  Plesser, Hans Ekkehard},
  title        = {NEST 3.9},
  month        = sep,
  year         = 2025,
  publisher    = {Zenodo},
  doi          = {10.5281/zenodo.17036827},
  url          = {https://doi.org/10.5281/zenodo.17036827},
}

@article {10.7554/eLife.47314,
  title={Brian 2, an intuitive and efficient neural simulator},
  author={Stimberg, Marcel and Brette, Romain and Goodman, Dan FM},
  journal={elife},
  volume={8},
  pages={e47314},
  year={2019},
  publisher={eLife Sciences Publications, Ltd}
}

@article{subramanian2021image,
  title={Image steganography: A review of the recent advances},
  author={Subramanian, Nandhini and Elharrouss, Omar and Al-Maadeed, Somaya and Bouridane, Ahmed},
  journal={IEEE access},
  volume={9},
  pages={23409--23423},
  year={2021},
  publisher={IEEE}
}

@ARTICLE{9187785,
  title={Digital steganography and watermarking for digital images: A review of current research directions},
  author={Evsutin, Oleg and Melman, Anna and Meshcheryakov, Roman},
  journal={IEEE Access},
  volume={8},
  pages={166589--166611},
  year={2020},
  publisher={IEEE}
}

@ARTICLE{9103079,
  title={Classifying melanoma skin lesions using convolutional spiking neural networks with unsupervised stdp learning rule},
  author={Zhou, Qian and Shi, Yan and Xu, Zhenghua and Qu, Ruowei and Xu, Guizhi},
  journal={IEEE Access},
  volume={8},
  pages={101309--101319},
  year={2020},
  publisher={IEEE}
}

@article{kim2020, 
  title={Spiking-yolo: spiking neural network for energy-efficient object detection},
  author={Kim, Seijoon and Park, Seongsik and Na, Byunggook and Yoon, Sungroh},
  booktitle={Proceedings of the AAAI conference on artificial intelligence},
  volume={34},
  number={07},
  pages={11270--11277},
  year={2020}
}

@inproceedings{lee2020spike,
  title={Spike-flownet: event-based optical flow estimation with energy-efficient hybrid neural networks},
  author={Lee, Chankyu and Kosta, Adarsh Kumar and Zhu, Alex Zihao and Chaney, Kenneth and Daniilidis, Kostas and Roy, Kaushik},
  booktitle={European Conference on Computer Vision},
  pages={366--382},
  year={2020},
  organization={Springer}
}

@manual{nest_iaf_psc_alpha,
  title        = {iaf\_psc\_alpha — NEST Simulator Documentation},
  author       = {{NEST Simulator Community}},
  year         = {2023},
  url          = {https://nest-simulator.readthedocs.io/en/v3.3/models/iaf_psc_alpha.html},
  note         = {Accessed: 2024-08-17},
}

@INPROCEEDINGS{8577655,
  title={A review on deep learning based image steganalysis},
  author={Tang, Yong-He and Jiang, Lie-Hui and He, Hong-Qi and Dong, Wei-Yu},
  booktitle={2018 IEEE 3rd Advanced Information Technology, Electronic and Automation Control Conference (IAEAC)},
  pages={1764--1770},
  year={2018},
  organization={IEEE}
}

@article{KARAMPIDIS2018217,
  title={A review of image steganalysis techniques for digital forensics},
  author={Karampidis, Konstantinos and Kavallieratou, Ergina and Papadourakis, Giorgos},
  journal={Journal of information security and applications},
  volume={40},
  pages={217--235},
  year={2018},
  publisher={Elsevier}
}

@INPROCEEDINGS{9340861,
  author={Stagsted, Rasmus Karnøe and Vitale, Antonio and Renner, Alpha and Larsen, Leon Bonde and Christensen, Anders Lyhne and Sandamirskaya, Yulia},
  booktitle={2020 IEEE/RSJ International Conference on Intelligent Robots and Systems (IROS)}, 
  title={Event-based PID controller fully realized in neuromorphic hardware: a one DoF study}, 
  year={2020},
  volume={},
  number={},
  pages={10939-10944},
  doi={10.1109/IROS45743.2020.9340861}
}

@article{MANDAL20221451,
    title = {Digital image steganography: A literature survey},
    journal = {Information Sciences},
    volume = {609},
    pages = {1451-1488},
    year = {2022},
    issn = {0020-0255},
    doi = {https://doi.org/10.1016/j.ins.2022.07.120},
    url = {https://www.sciencedirect.com/science/article/pii/S002002552200809X},
    author = {Pratap Chandra Mandal and Imon Mukherjee and Goutam Paul and B.N. Chatterji}
}

@ARTICLE{9599677,
  author={Gurunath, R. and Klaib, Mohammad Fadel Jamil and Samanta, Debabrata and Khan, Mohammad Zubair},
  journal={IEEE Access}, 
  title={Social Media and Steganography: Use, Risks and Current Status}, 
  year={2021},
  volume={9},
  number={},
  pages={153656-153665},
  doi={10.1109/ACCESS.2021.3125128}
}

@article{cisska-lsb,
    author = {Muhammad, Khan and Ahmad, Jamil and Rehman, Naeem Ur and Jan, Zahoor and Sajjad, Muhammad},
    title = {CISSKA-LSB: color image steganography using stego key-directed adaptive LSB substitution method},
    year = {2017},
    issue_date = {March     2017},
    publisher = {Kluwer Academic Publishers},
    address = {USA},
    volume = {76},
    number = {6},
    issn = {1380-7501},
    url = {https://doi.org/10.1007/s11042-016-3383-5},
    doi = {10.1007/s11042-016-3383-5},
    journal = {Multimedia Tools Appl.},
    month = mar,
    pages = {8597–8626},
    numpages = {30}
    }

@article{solak2019image,
  title={Image steganography based on LSB substitution and encryption method: adaptive LSB+ 3},
  author={Solak, Serdar and Alt{\i}n{\i}{\c{s}}{\i}k, Umut},
  journal={Journal of Electronic Imaging},
  volume={28},
  number={4},
  pages={043025--043025},
  year={2019},
  publisher={Society of Photo-Optical Instrumentation Engineers}
}

@article{zhang2019steganogan,
  title={SteganoGAN: High Capacity Image Steganography with GANs},
  author={Zhang, Kevin Alex and Cuesta-Infante, Alfredo and Veeramachaneni, Kalyan},
  journal={arXiv preprint arXiv:1901.03892},
  year={2019},
  url={https://arxiv.org/abs/1901.03892}
}

@article{Aletheia,
  author = {Lerch-Hostalot, Daniel and Megías, David},
  doi = {10.21105/joss.05982},
  journal = {Journal of Open Source Software},
  month = jan,
  number = {93},
  pages = {5982},
  title = {{Aletheia: an open-source toolbox for steganalysis}},
  url = {https://joss.theoj.org/papers/10.21105/joss.05982},
  volume = {9},
  year = {2024}
}

@InProceedings{DIV2K2017,
	author = {Agustsson, Eirikur and Timofte, Radu},
	title = {NTIRE 2017 Challenge on Single Image Super-Resolution: Dataset and Study},
	booktitle = {The IEEE Conference on Computer Vision and Pattern Recognition (CVPR) Workshops},
	month = {July},
	year = {2017}
}

@article{steganogan,
  title={SteganoGAN: High Capacity Image Steganography with GANs},
  author={Zhang, Kevin Alex and Cuesta-Infante, Alfredo and Veeramachaneni, Kalyan},
  journal={arXiv preprint arXiv:1901.03892},
  year={2019},
  url={https://arxiv.org/abs/1901.03892}
}

@inproceedings{neeta2006implementation,
  title={Implementation of LSB steganography and its evaluation for various bits},
  author={Neeta, Deshpande and Snehal, Kamalapur and Jacobs, Daisy},
  booktitle={2006 1st international conference on digital information management},
  pages={173--178},
  year={2006},
  organization={IEEE}
}

\end{document}